\documentclass{article}
\usepackage{hyperref} 
\usepackage{url}
\usepackage{amscd,amsmath,amsthm,amssymb}
\usepackage{enumerate,varioref}
\usepackage{epsfig}
\usepackage{graphicx}
\usepackage{subfig}
\usepackage{mathtools}
\usepackage{tikz}
\usepackage{algorithmicx}
\usepackage{algorithm}
\usepackage{algpseudocode}
\usepackage{xcolor}

\theoremstyle{definition}

\theoremstyle{remark}

\newtheorem*{rem}{Remark}

\title{On Near Optimal Spectral Expander Graphs of Fixed Size}
\author{Clark Alexander\\ email: \href{mailto:gcalexander1981@gmail.com}{the author}}

\begin{document}
	
	\maketitle
	
	\begin{abstract}
		We present two powerful heuristic methods for building random regular graphs and finding a near optimal spectral gap in a finite regular graph.
	\end{abstract}

	\tableofcontents
	
	\section{Introduction}
	From the 1980s and 1990s one has seen several methods for generating random regular graphs [MW,W].  While generating a random graph is extremely easy, the constraint of requiring regularity is what produces the difficulty.  There is, however, some renewed interest in random regular graphs as we see that many random regular graphs become near spectral expanders [BB].  In addition, regular graphs allow for consistent edge labeling in a rotation map. [A2] which allows one to efficiently compute a discrete time quantum walk [A1].  
	
	This work is presented in two major parts.  The first is presenting an extremely simple method for producing a random regular graph with a fixed number of vertices and degree of regularity. This method involves a constructive method for building a regular graph of fixed size and a randomization algorithm by which we can move around a fixed number of edges.  
	
	The second part which entails the lion's share of the present work shows how to produce a Metropolis Coupled Simulated Annealing algorithm for near global optimization of spectral gaps in random regular graphs of fixed size.  In the initial stages of this work, we had tried to use a simple simulated annealer, which in large part was able to find Ramanujan graphs quickly and easily.  However, during our computational explorations, we found that the method we used for picking a ``neighboring" solution greatly affected the overall quality of the solution.  Furthermore, switching the method of choosing a neighbor during the annealing process again affected the quality of solution.  Thus the next level of complexity called for coupling the annealing solutions.  This draws us back to a Metropolis Coupled Markov Chain Monte Carlo (MCMCMC) [MNL] which entered into the computational space cerca 1997 as a way of having a more effective Monte Carlo method for multi-modal distributions.  We have modified the MCMCMC to be a Metropolis coupled simulated annealer (MCSA) to look for near optimal spectral expander graphs. We give some numerical results and some empirical evidence in support of an open question as to whether there are infinitely many 7-regular Ramanujan graphs.  In short, we have tested every graph with on even number of vertices from 8 to 1000 vertices, and have always been able to produce a Ramanujan graph within a matter of seconds. We have also tested graphs of size 1000 to 2000 adding 20 vertices at a time and have again been able to find Ramanujan graphs, but the computational effort grows substantially after approximately 1200 vertices.  Nonetheless, we have found several 7-regular 2000 vertex Ramanujan graphs and the adjacency matrix for one such is available upon request from the author.

    \section{Part 1: The Algorithms for Randomization}
    \subsection{Constructing a Regular Graph of Fixed Size}
    In order to draw a regular graph we only need to check one criterion, that is, the product of number of vertices with degree of regularity is even.  In our case we will only consider simple undirected graphs with no loops.  From our first week of graph theory class we remember for a graph $G = (V,E)$ with degree of regularity $d$ we have.
    \begin{equation}
    \label{even} deg(G) = 2|E| = d|V|
    \end{equation}
    
    We cannot satisfy this if $d$ and $|V|$ are both odd.
    
    Thus we can split our regular graph generation into two cases for the parity of $d$.  In the first, and easier case, we let $d$ be even.  This allows us to pick any number of vertices (greater than $d$).
    The first step is to label our vertices $1$ to $|V|$ ( or 0 to $|V|-1$ mod $|V|$).  And create the cycle graph $C_{|V|}$.  This gives us a 2 regular graph.  Now we iterate for $i=2$ to $d/2$ connecting $v_n$ to $v_{n+i}$.  Below is pseudo which has been implemented in both Python(3) and Julia (1.4).  This is a pseudo-code for generating the adjacency matrix of a regular graph. We assume that the product of vertices and degree of regularity is even.
    
	\begin{algorithm}
	\caption{Rotation map from the adjacency matrix; Pseudo-code}
	\label{Adjacency of Regular Graph}
	\begin{algorithmic}[1]
		\Procedure{GenerateRegularGraph}{vertices, degree)}
			\State $n$ = vertices
			\State $d$ = degree
		\If{$d \mod 2 = 1$}
			\State $A = eye(n, k = n//2)$
			\Comment This corresponds to connecting diametrically opposed vertices.
			\For{ $L$ in $1 : d//2$}
				\State $A += eye(n, k = L+1)$
				\State $A += eye(n, k = n-(L+1))$
			\EndFor
			\State $A = (A + A^{t})$
			\Comment Symmetrize the matrix
		\Else
			\Comment We don't have diametrically connected vertices
			\State $A = zeros(n,n)$
			\For{ $L$ in $1 : d//2$}
				\State $A += eye(n, k = L+1)$
				\State $A += eye(n, k = n-(L+1))$
			\EndFor	
			\State $A = (A + A^{t})$
			\EndIf
		\State return $A$   
		\EndProcedure
	\end{algorithmic}	
    \end{algorithm}

    Let's take a look at three potential examples.  These are all produced by the networkx package in Python. Figure 1 gives 4 regular graphs with no randomness the number of vertices and degrees of regularity are $(9,4), (12.4), (12,6), (14,5)$ respectively.
    
    \begin{figure}[!ht]
    	\centering
    	\subfloat[9 vertex, 4-regular]{{\includegraphics[width=5.5cm]{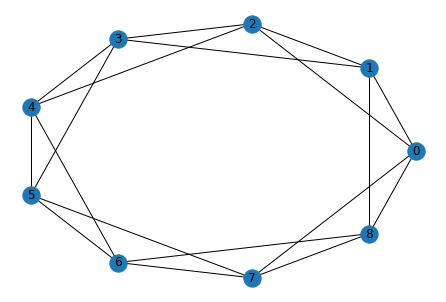} }}%
    	\qquad
    	\subfloat[12 vertex 4-regular]{{\includegraphics[width=5.5cm]{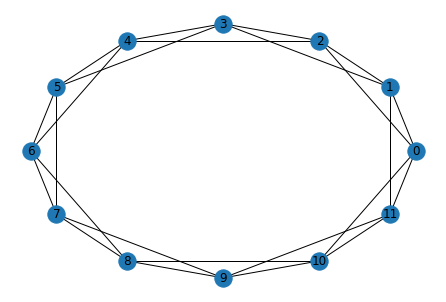} }}%
    	\newline
    	\subfloat[12 vertex 6-regular]{{\includegraphics[width=5.5cm]{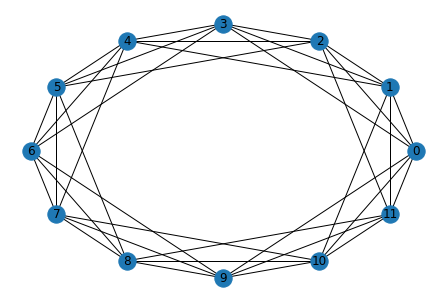} }}%
    	\qquad
    	\subfloat[14 vertex 5-regular]{{\includegraphics[width=5.5cm]{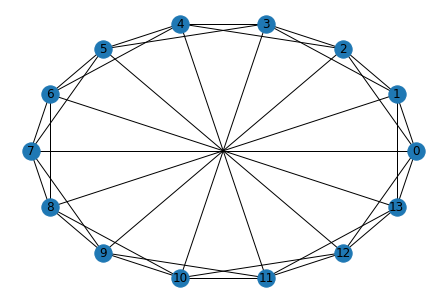} }}%
    	%\caption{}%
    	\caption{Regular graphs with no randomness.}
    	\label{fig:Constructed Regular Graphs}
    \end{figure}

    \subsection{The Algorithm for Randomization}
    
    The general algorithm is quite straight forward once we have a regular graph in place.  In plain English we algorithm looks something like:
    \begin{enumerate}
    	\item[1.] Pick two edges at random, swap one end point of each.
    	\item[2a.] If the graph is still regular with the same degree and there are no loops accept the switch.
    	\item[2b.] If not, pick again.
    	\item[3.] Repeat until ``sufficiently" randomized.
    \end{enumerate}
    
    The supporting procedures of checking for loops (i.e. the diagonal of the adjacency matrixhas a nonzero element) and that the graph is still regular (all rows sum to the same degree) will be assumed.  Now we come to the pseudo-code implementation of the graph randomization algorithm. First, we need the ability to switch edges.  In the adjacency matrix representation the edges are represented by locations in the matrix which are 1.  So the procedure \emph{getEdges} is simply find the nonzero locations in the adjacency matrix
    
    	\begin{algorithm}
    	\caption{Switch Edges; Pseudo-code}
    	\label{Switch Edges}
    	\begin{algorithmic}[1]
    		\Procedure{SwitchEdges}{vertices, degree}
    		\State $A\gets GenerateRegularGraph$(vertices, degree)
    		\State $B \gets A$ 
    		\Comment In python we need $A$.copy() to get a different memory pointer
    		\State edges $\gets$ getEdges($A$)
    		\State Randomly choose two edges $e1, e2$ from edges.
    	    \State $new1 = [e1[1],e2[2]]$
    	    \State $new2 = [e2[1], e1[2]]$
    	    \State $B[e1[1],e1[2]] = 0$
    	    \State $B[e2[1],e2[2]] = 0$
    	    \State $B[e1[2],e1[1]] = 0$
    	    \State $B[e2[2],e2[1]] = 0$
    	    \State $B[new1[1],new1[2]] = 1$
    	    \State $B[new1[2],new1[1]] = 1$
    	    \State $B[new2[1],new2[2]] = 1$
    	    \State $B[new2[2],new2[1]] = 1$
    		\If{is\_regular($B$) and np\_loops($B$)}
    		    \State return $B$   
    		\Else
    			\State return $A$    
    		\EndIf	
    		\EndProcedure
    	\end{algorithmic}	
    \end{algorithm}   
    
    \begin{algorithm}
    	\caption{Generate Random Regular graph; Pseudo-code}
    	\label{Random Graph}
    	\begin{algorithmic}[1]
    		\Procedure{GenerateRandomGraph}{vertices, degree, switches}
    		\State $A\gets GenerateRegularGraph$(vertices, degree)
    		\State $B \gets$ SwitchEdges$(A)$ 
    		\For{ $k$ in 1:switches}
    			\State $B \gets$ SwitchEdges$(B)$
    		\EndFor	
    		\State return $B$	
    		\EndProcedure
    	\end{algorithmic}	
    \end{algorithm}

    \subsection{Examples of Randomized Regular Graphs}
    
    Now we give three pairs of examples.  In order to show that we can use this to generate random graphs of nearly arbitrary degree.  We'll give examples of size (20,8), (30,13), and the much larger (500,25).

    \begin{figure}[!ht]
    	\centering
    	\subfloat[20 vertex 8-regular nonrandom]{{\includegraphics[width=4.9cm]{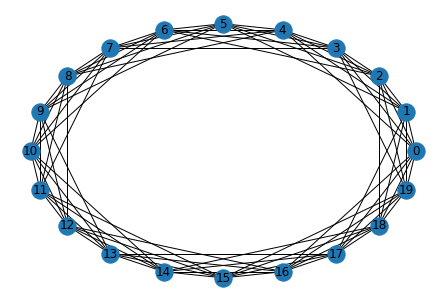} }}%
    	\qquad
    	\subfloat[20 vertex 8-regular random]{{\includegraphics[width=5.5cm]{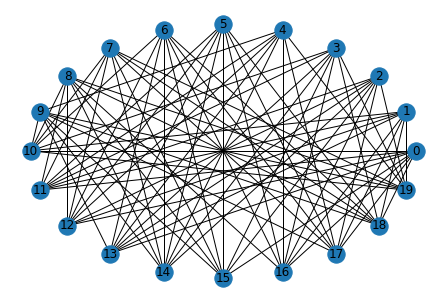} }}%
    	\caption{20 vertex 8-regular graphs}%
    	\label{fig:20-8}%
    \end{figure}

    \begin{figure}[!ht]
    	\centering
    	\subfloat[30 vertex 13-regular nonrandom]{{\includegraphics[width=4.9cm]{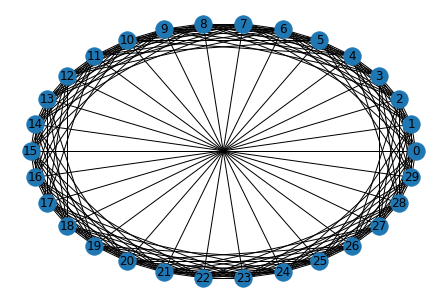} }}%
    	\qquad
    	\subfloat[30 vertex 13-regular random]{{\includegraphics[width=5.5cm]{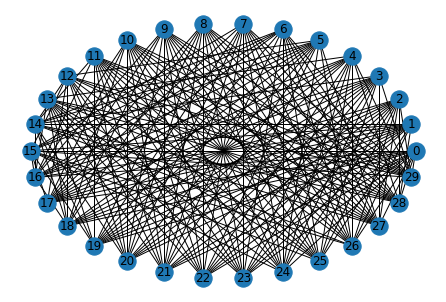} }}%
    	\caption{30 vertex 13-regular graphs}%
    	\label{fig:30-13}%
    \end{figure}
    
    For the larger example, the plots are not so enligthening, but we shall show them here and give timing so that we can demonstrate the overall efficacy of our ability to randomize graphs of nearly arbitrary degree.
    
     \begin{figure}[!ht]
    	\centering
    	\subfloat[500 vertex 25-regular nonrandom]{{\includegraphics[width=4.9cm]{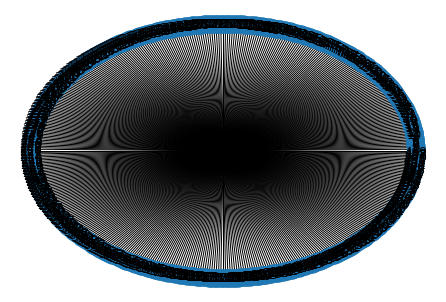} }}%
    	\qquad
    	\subfloat[500 vertex 25-regular random]{{\includegraphics[width=5.5cm]{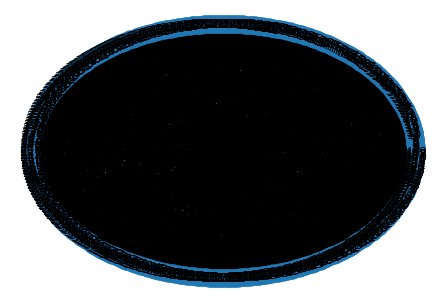} }}%
    	\caption{500 vertex 25-regular graphs}%
    	\label{fig:500-25}%
    \end{figure}
    
    The second graph in \ref{fig:500-25} is the results of 1500 random edge switches and on the author's laptop took 53.6s to complete.
    
    And for posterity let's examine the spectral propertires of these two graphs.
    
    \begin{figure}[!ht]
    	\centering
    	\subfloat[500 vertex 25-regular nonrandom]{{\includegraphics[width=4.9cm]{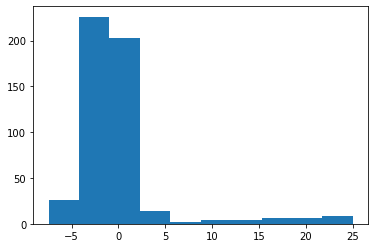} }}%
    	\qquad
    	\subfloat[500 vertex 25-regular random]{{\includegraphics[width=5.5cm]{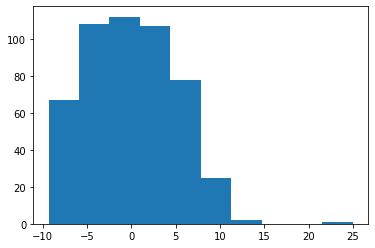} }}%
    	\caption{500 vertex 25-regular graphs eigenvalue histograms}%
    	\label{fig:500-25eigenvalues}%
    \end{figure}
    
    It's clear from the spectrum that both graphs are 25 regular and that the randomized graph is closer to the semi-circular distribution than the nonrandom graph.

    \section{Part 2: The Search for Ramanujan Graphs which do not have Explicit Constructions}
    \subsection{How to Anneal a Regular Graph}
    
    By the construction in \S2.1 we achieve a regular graph of fixed size.  In particular we build the adjacency matrix of a regular graph of fixed size.  This construction, however, has a small spectral gap and a large graph diameter.  Using our randomization scheme we wish to construct an adjacency matrix of a regular graph with a larger spectral gap. We have seen before [BB] that a random regular graph comes close to a Ramanujan bound for spectral expansion.  The goal at present is to use our randomization scheme to search for ever larger spectral gaps.  In reality, we will use a simulated annealer to minimize the normalized $\lambda_2$. Looking a little further down \ref{fig:Eigenvalues by random switches} shows how eigenvalues perform with single random switches.

    We see that that randomization scheme first brings $\lambda_2$ to within the Ramanujan threshold quickly, but eventually moves above it.  This does not, however, get to the weak bound of Alon-Bopanna, nor the strong bound.  It is our goal to see how close we can get to these bounds.  The striking feature or many graphs is how quickly we can cross the simple Ramanujan threshold.  
    
    The general structure of a simulated annealing algorithm is straight-forward:
    \begin{enumerate}
    	\item Guess a solution at random
    	\item Score that solution
    	\item Pick a ``neighbor" solution
    	\item Score the neighbor.
    	\item If the neighbor is better, make it the new solution
    	\item if neighbor is worse, still accept neighbor if not too far off
    	\item Lower the algorithm's ``temperature"
    	\item Repeat until satisfied
    \end{enumerate}
    
    This is a gross oversimplification, but there are two salient points we need to consider.  The first being: How do we pick an effective ``neighbor" solution.  The second being: How do we deal with the temperature parameter?  
    Changes in either of this points can drastically affect the quality of solutions in a simulated annealer.  In particular, lowering the temperature too quickly (i.e. quenching) tends to land a solution in the first deep local minimum the algorithm finds.  On the other hand, lowering the temperature too slowly, takes too long and begins to look like a brute-force solution, which we don't have time for.  
    
    In the case of the random regular graph, we will set our annealing algorithm to look for the lowest possible normalized $\lambda_2$ rather than the largest spectral gap.  As it is set up, however, we will never achieve a bipartite graph as the second normalized eigenvalue is 1. In particular a bipartite graph of degree $d$ has eigenvalues $\pm d$, which we normalize and consider absolute values.  Thus this algorithm ignores bipartite graphs as good expanders since we wish to achieve the steady state vector quickly by starting a random walk at a single vertex rather than in a superposition of vertices in different partition sets.  We see, from the figure \ref{fig:Eigenvalues by random switches} that switching a pair of edges does well, however, for larger and sparser graphs, this takes too long. Consider, for example, a large cubic or 4-regular graph. That is a graph with several thousand vertices.  A single edge switch does very little to affect the eigenvalue gap.  Thus, one may consider using multiple edge swaps as the neighboring solution.  This means that a ``neighbor" graph should swap a dozen or so pairs of edges before measuring $\lambda_2$.
    
    As with a standard annealing algorithm we want to move from essentially a random walk when the temperature is hot to a stochastic gradient descent type algorithm.  We, however, wish to slow down not only the random selection of temporary solutions, but also slow down or speed up the edge switching as necessary.  Thus we have decided to move from a simple single simulated annealer to a coupled annealer.
      
    \subsection{Metropolis Coupled Simulated Annealing}
    
    A single simulated annealer is an example of a Markov chain Monte Carlo method tuned for optimization as opposed to finding a posterior probability distribution.  As discussed in the previous section we wish to couple multiple simulated annealing algorithms to find a solution which is closerto a global maximum.  The idea of coupling was first (published) in the late 1990s [MNL] with an eye toward bimodal and multimodal distributions where MCMC algorithms are slow to converge.  From the perspective of this work, we shall leverage multiple optimizers to look for a (closer to) global optimum.  In order to tune our coupling we implement the following procedure:
    \begin{enumerate}
    	\item Initialize $N$ simulated annealers with initial temperatures $\tau_1,\dots,\tau_N = 1$ with different cooling rates $\alpha_1\dots,\alpha_N$ These are called chains.
    	\item Each chain gets a different method of selecting a neighboring solution.  For this work the $n^{th}$ chain has a neighbor selection of $n$ random edge switches.
    	\item For each chain run one step of the simulated annealing algorithm and cool each chain $\tau_i \gets \tau_i *\alpha_i$
    	\item Rank order the chains by best solution.  The best current solution gets the lowest temperature.  This is the ``cold chain." 
    	\item Randomly select two other chains, score them and compute an acceptance probability for switching (or more simply one can simply exchange them without the extra computation).  
    	\item Select a number from a uniform variable $U \sim [0,1]$.  If the acceptance probabilty is higher than the randomly drawn number, switch the chains.  That is, switch their neighbor solutions, temperatures and cooling rates.
    	\item Repeat steps 3-6 until the hottest chain has reached the temperature threshold, or we have reached an acceptable bound (e.g. We have found a Ramanujan graph, or reached a weak lower bound for $\lambda_2$).
    \end{enumerate}

    \subsection{Some Bounds and Asymptotics}
    
    For this section we will present the results as they were originally stated and then show the results for normalized eigenvalues.  We will also let a graph $G = (V,E)$ have degree of regularity $d$ throughout.

    \subsubsection{Eigenvalues}
    
    The original bound for an expander graph to be a Ramanujan graph is 
    \[
    |\lambda_2| < 2\sqrt{d-1}
    \]
    In our case, of course, this means
    \[
    |\lambda_2| < \frac{2\sqrt{d-1}}{d}
    \]
    Additionally from \cite{N,W1} we have several other bounds.  There is a weak bound in which we have
    \begin{equation}
    |\lambda_2|\ge \frac{2\sqrt{d-1}}{d}\left(1 - \frac{1}{2r}\right)
    \end{equation}
    where $r\in \mathbb{N}$.  This particular $r$ is chosen so that $k \le r-1$ where 
    \[
    k = \log_{\sqrt{d-1}}(|V|)
    \]
    The basic construction comes from finding an eigenvector corresponding to a particular vertex within $G$.  With a small inspection one can see that as $|V| \rightarrow \infty$ this bound becomes the original Ramanujan bound.  
    In our case, we will look for eigenvalues which cross the (slightly) higher bound
    \[
    |\lambda_2| \le \frac{2\sqrt{d-1}}{d}\left(1 - \frac{1}{2d}\right) 
    \]
    We will demonstrate that for smaller graphs(up to roughly 2000 vertices) we can often get below this bound.
    
    There is one additional much stronger bound
    \begin{equation}
    |\lambda_2| \ge \frac{2\sqrt{d-1}}{d} - \frac{2\sqrt{d-1} - 1}{d\lfloor m/2 \rfloor}
    \end{equation}
    where $m$ is the diameter of the graph.  In [N] this is a strict lower bound.  Again, one can see as $|V|\rightarrow \infty$ we again approach the Ramanujan bound.  This bound as we will show empirically can not be crossed, and even approaching this bound is difficult except for complete graphs or particularly tiny graphs.  For example, we have enough computing fire power that we can check every regular graph of size $|V| \le 14$.
    
    \begin{rem}
    	We can precompute the value of the diameter ($m$) of the graph by a simple counting argument.
    	This is done by assuming we have a $d$-ary tree.  The distance from the root to any leaf is the radius which is half the diameter.
    	\begin{equation}
    	r = \frac{m}{2} \ge \lceil\log_{d}(|V|)\rceil
    	\end{equation}
    \end{rem}

    \subsubsection{The Number of Regular Graphs}
    For smaller graphs we have exact calculations of the number of regular graphs \cite{OEIS1, OEIS2}.  We also have in \cite{MW} (p58 theorem 1) the number of $d$-regular graphs on $n$ vertices is 
    \begin{equation}
    |\mathcal{G}_{n,d}| \sim \frac{(nd)! \exp(1- d^2/4)}{(nd/2)! 2^{nd/2}}\left(1 + o(1)\right)
    \end{equation}
    
    Rearranging this a bit we get the logarithmic count of $d$-regular graphs on $n$ vertices is
    \begin{equation}
    \log(|\mathcal{G}_{n,d}|) \sim \frac{nd}{2}(\log(nd) - 1) + \frac{\log(2)}{2} + \frac{(1-d^2)}{4}
    \end{equation}
    
    \subsubsection{Cycles in Random Regular Graphs}
    We see in \cite{BB, W2} that the number of cycles of length $k$ in a $d$-regular random graph is asymptotically a Poisson random variable $X_k \sim$Pois$(\lambda_k)$ with mean
    \begin{equation}
    \lambda_k = \frac{(d-1)^k}{2k} 
    \end{equation}
    
    \subsection{Known Constructions}
    There are several well-known constructions for families of Ramanujan graphs, in particular, LPS, Morgenstern, Paley, MSS, Zig-Zag, and Super-Singular Isogeny graphs.  Additionally [F,BB] points out that with high probability a $d$-regular random graph is Ramanujan or near Ramanujan (ie $\lambda_2 \le 2\sqrt{d-1} + \varepsilon$)
    
    The known constructions cover the following cases
    %table of constructions here
    \[
    \begin{tabular}{|p{3.5cm} | c | p{3.5cm} |}
    \hline
    Construction & Degree of Regularity & Conditions \\
    \hline
    LPS \cite{LPS} & $p+1$ & $p\equiv 1 \mod4$\\
    \hline
    Margulis-Gabber-Galil \cite{Mar,GG} & 4 & $n^2$ vertices or infinite\\
    \hline
    Morgenstern \cite{Mor} & $q+1$ & $q = p^n$ for $p$ prime\\
    \hline
    Zig-Zag \cite{RVW} & $d_2^2$ & $d_1, d_2$ are degrees of regularity of the two graphs to be multiplied\\
    \hline
    Paley \cite{Pa, ER}& $\frac{q^n - 1}{2}$ or $\frac{p^{2n}-1}{2}$ & $q \equiv 1 \mod 4$ prime, $p \equiv 3 \mod 4$ prime\\
    \hline
    SuperSingular Isogeny Graphs \cite{Pi, EHLMP} & $\ell + 1$ & $\ell$ prime.\\
    \hline
    Bipartite Graphs \cite{MPS} & any & Graph must be Bipartite\\
    \hline
    \end{tabular}
    \]
    
    We see now that the smallest positive integer for which a family of constructions is not known is $d=7$ for graphs with chromatic number 3 or more.  We also have a few other small unknown cases $d = 11,13,15,19,21,22,23,24$  Additionally, we see that there are some graphs with specific numbers of vertices which are not known in constructions.  For example, the Zig-Zag product requires the number of vertices to be composite and Paley graphs have vertices which are elements of a finite field so we have powers of primes.  In the case where $p\equiv 1\mod 4$ the smallest degree of regularity for a Paley graph is $(p-1)/2$ which quickly becomes quite large.  We will explore some relatively smaller graphs for visualization purposes and some moderate sized graphs for the purposes of numerical demonstration.
    
    \begin{rem}
    	The methods of [Mar,GG] cover 4-regular graphs, but with a perfect square number of vertices.  The MCSA finds a 17 vertex Ramanujan graph which is 4-regular in $\sim 0.17 s$ using a flag to stop at finding such a graph.  Additionally, the MCSA finds a near optimal graph of 17 vertices ($\lambda_2 \approx 0.6204$) in $\sim 53.45s$.  The Ramanujan threshold for a 4-regular graph is $\approx 0.866$.  Among 17 vertex 4-regular graphs there are $\approx 1.15 \times 10^{46}$ according to [MW]. This is a small demonstration of the efficacy of MCSA.  In the next section we shall explore some numerical results more deeply.
    \end{rem}

    \subsection{Some Numerical Results}
    
    To demonstrate the efficacy of the heuristic search involved, let's begin with some well-known graphs which are near optimal, the fullerenes [AKS].  In particular we'll look explicitly at the dodecahedron and the ``bucky-ball." These are well known 3-regular (cubic) graphs with 20 and 60 vertices respectively. To get our eyes adjusted to the way we will draw graphs, here are depictions of the respective fullerene graphs.
    \begin{figure}[!ht]
    	\centering
    	\subfloat{{\includegraphics[width=5.5cm]{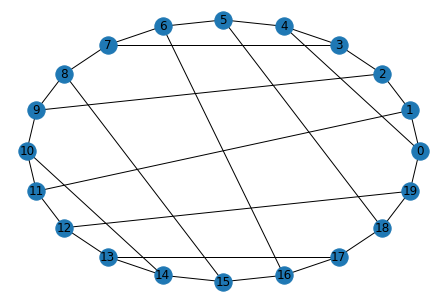}}}
    	\subfloat{{\includegraphics[width=5.5cm]{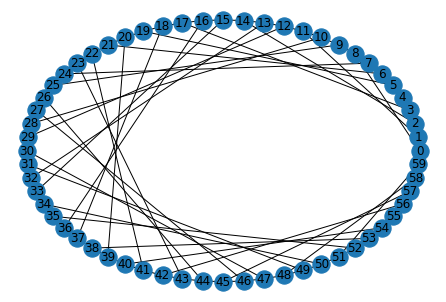}}}
    	\caption{The Dodecahedron and the Bucky Ball}
    \end{figure}
    
    In our scenario we initialize two graphs, a 20 vertex 3-regular and a 60 vertex 3-regular which look like this.
    \begin{figure}[!ht]
    	\centering
    	\subfloat{{\includegraphics[width=5.5cm]{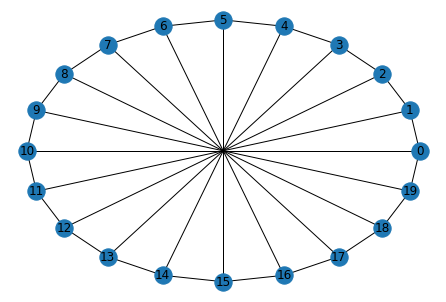}}}
    	\subfloat{{\includegraphics[width=5.5cm]{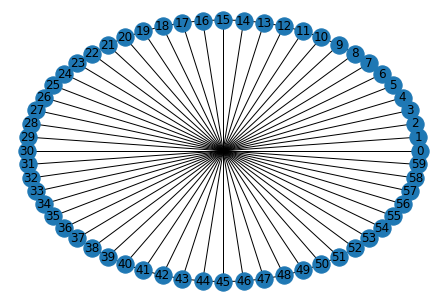}}}
    	\caption{3-regular graphs from explicit construction}
    \end{figure}
     
    These graphs are nowhere near expanders, much less optimal expanders. In particular the normalized threshold for a cubic graph to Ramanujan is $\lambda_2 \le 2\sqrt{2}/3 \approx 0.9428$.  In our case the initialized graphs have $\lambda_2 \approx 0.96737$ for the 20 vertex graph and  $\lambda_2 \approx 0.99634$ for the 60 vertex graph.       
     
    Again, from [F,BB] we know with high probability that a random regular graph will be Ramanujan.  So let's 
    follow two progressions randomly switching edges for each of these graphs and see how often they fall below the required threshold.
    
    \begin{figure}[!ht]
    	\centering
    	\subfloat{{\includegraphics[width=6.5cm]{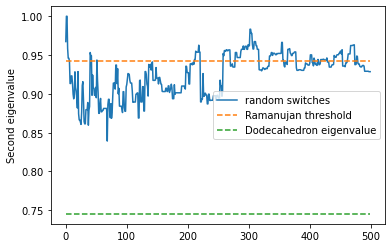}}}
    	\subfloat{{\includegraphics[width=6.5cm]{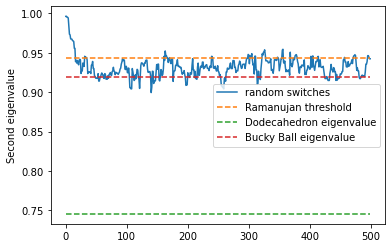}}}
    	\caption{Eigenvalues of 20 and 60 vertex graphs via random switching}
    		\label{fig:Eigenvalues by random switches}%
    \end{figure}
    
    A couple of immediate observations.  The smaller the graph, the easier it is to be a spectral expander.  This agrees with our intuition about edge expanders as we require very few moves with a small number of vertices to reach any vertex via random walk when the graph is ``small."  Larger graphs, however, require more moves and we see this in the fact that a single random switch affects $\lambda_2$ far less in the 60 vertex case than in the 20 vertex case.  We also see that a vast majority of random regular graphs are below the Ramanujan threshold.  In particular in two random trials of 499 graphs, we have 345 such graphs below the threshold in the 20 vertex case and 432 in the 60 vertex case.  Additionally we have 68 and 196 graphs below the bucky ball line respectively.  However, nothing gets below the dodecahedron line.
    
    Consider now two graphs that we have generated with our Metropolis coupled simulated annealer.  These are noth first run algorithms, and thus if we had tuned our parameters very specifically, we could have achieved lower eigenvalues.  
    
        \begin{figure}[!ht]
    	\centering
    	\subfloat[$\lambda_2 \approx 0.76759$]{{\includegraphics[width=6.5cm]{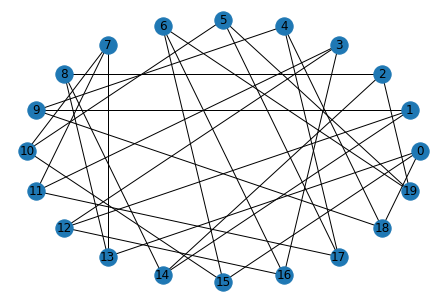}}}
    	\subfloat[$\lambda_2 \approx 0.8576$]{{\includegraphics[width=6.5cm]{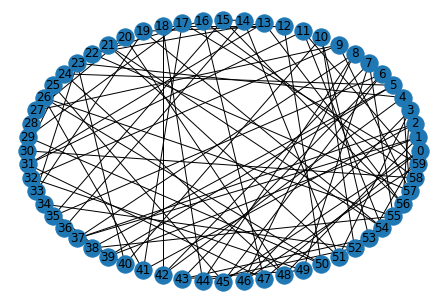}}}
    	\caption{Expanders of 20 and 60 vertex graphs via MCSA}
    		\label{fig:Near Optimal 20 and 60 vertices}%
        \end{figure}
    
    Again we notice that neither of these graphs can quite hit the dodecahedon threshold, but they have both surpassed the bucky ball threshold easily as well as exceeding any graph achieved by single random switches.
    
    Now for just a moment let's turn our attention to 7-regular graphs.  This is one of the cases which is not explicitly covered in known constructions. Drawing a graph with more than 60 or so vertices becomes rather unattractive on a page this small, so we will stick to 50 vertices for the sake of demonstration.  In a timed run, we flagged the algorithm to stop once it had a achieved a Ramanujan graph.  This particular 50 vertex, 7-regular graph cost the author's laptop 1.1373 seconds of cpu time.  The normalized threshold for a 7-regular graph is 
    \[
    \lambda_2 = \frac{2\sqrt{6}}{7} \approx 0.69985
    \]
    
   \begin{figure}[!ht]
   	\centering
   	\includegraphics[width=0.95\textwidth]{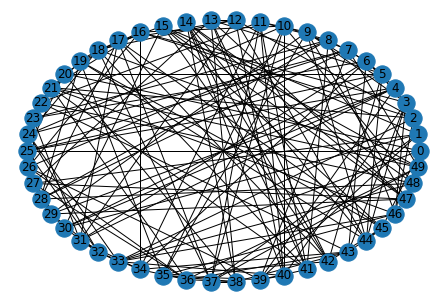}
   	\caption{A quick Ramanujan graph with $\lambda_2 \approx 0.598987$}
    \end{figure}
    
    As we can see, this is significantly lower than the required Ramanujan threshold.  This is even below a weak optimal expander as defined in Equation 2.

    Now consider a graph that we did not flag.
    
       \begin{figure}[!ht]
    	\centering
    	\includegraphics[width=0.95\textwidth]{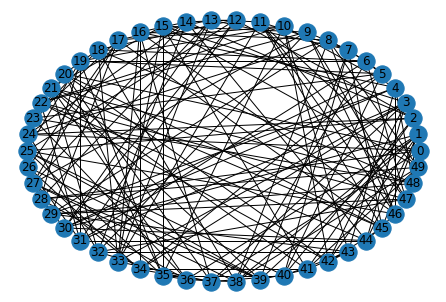}
    	\caption{A near optimal graph with $\lambda_2 \approx 0.5374037$}
        \end{figure}
    
    In the case of 7-regular graphs the thresholds we're dealing with are 
    \[
    \begin{tabular}{l | c}
    	Ramanujan & $\approx 0.69985$\\
    	\hline 
    	weak optimal & $\approx 0.649865$\\
    	\hline 
    	strict lower bound & $\frac{1}{7} \approx  0.142857$
    \end{tabular}
    \]
    
    In nearly every case attempted by the author the heuristic optimizer was able to find a graph close to (and in many cases below) the weak optimal expansion constant. In every case, we were able to find a Ramanujan graph, but again, this is not surprising based on the work of [F]  However, we were never able to find a graph which met the strict lower bound as given by [N].  In simple numerical trials we were able to find Ramanujan graphs with degree of regularity 7 for every (even) number of vertices between 8 and 2000. This work, however, does not answer the question of whether there is an explicit construction for such graphs.
    
    While the drawing of a graph with 2000 vertices is not very attractive, we can still show the histogram of normalized eigenvalues to show the spectral expansion.  Below is a histogram of a 2000 vertex 7-regular graph discovered by Metropolis coupled simulated annealing.  The author has also added a spreadsheet of its adjacency matrix into github (email author for address) so that researchers can explore its random properties further at their desire.
    
    \begin{figure}[!ht]
    	\centering
    	\includegraphics[width=0.95\textwidth]{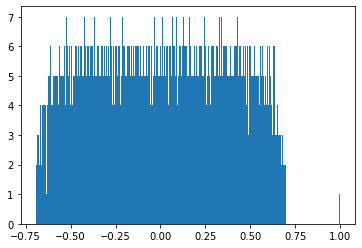}
    	\caption{Histogram of 2000 vertex 7-regular Ramanujan graph normalized eigenvalues; $\lambda_2 \approx 0.69407$}
    \end{figure}    
    
    \section*{Appendix: Additional Algorithms}

    	\begin{algorithm}
    	\caption{Partial Annealing}
    	\label{Partial Annealing}
    	\begin{algorithmic}[1]
    		\Procedure{PartialAnneal}{matrix,  trials, temperature, cooling\_rate, switches}
    	    \State $A$ = matrix
    	    \State $T$ = temperature
    	    \State $\lambda0$ = computeSecondEigenvalue$(A)$
    	    \State solution0 = A
          	\State best\_$\lambda$ = $\lambda0$
    	    \State best\_solution = solution0
    	    \For{k in 1:trials}
            \State	solution1 = getNeighbor(solution0, switches)
    	    \State $\lambda1$ = computeSecondEigenvalue(solution1)
    		\If {$\lambda1 < $ best\_$\lambda$}
    		\State best\_$\lambda$ = $\lambda1$
    		\State best\_solution = solution1 \Comment{Keep the global best in storage}
    		\EndIf
    		\If {$\lambda1 < \lambda0$}
    		\State solution0 = solution1
    		\State $\lambda0$ = $\lambda1$
    		\Else
    		\State ap = acceptance\_probability($\lambda0$, $\lambda1$, $T$)
    		\EndIf
    		\If {ap $>$ rand()}
    		\State solution0 = solution1
    		\State $\lambda0 = \lambda1$
    		\EndIf
    		\EndFor
    	\State $T *= $cooling\_rate
    	\State return best\_solution, best\_$\lambda$, $T$
    	\EndProcedure
    	\end{algorithmic}	
    \end{algorithm}   
    
    	\begin{algorithm}
    	\caption{Define Coupling}
    	\label{Define Coupling}
    	\begin{algorithmic}[1]
    		\Procedure{DefineCoupling}{(vertices, degree, chains, min\_cooling, max\_cooling)}
    		\State graphs = [generateRegularGraph(vertices, degree) for k in range(chains)]
    		\State initial\_temperatures = ones(chains)
    		\State cooling\_rates = [start = min\_cooling, end = max\_cooling, spacing = (max\_cooling - min\_cooling) / chains]
    		\State return graphs, initial\_temperatures, cooling\_rates
    		\EndProcedure
    	\end{algorithmic}	
    \end{algorithm}   
    
    	\begin{algorithm}
    	\caption{Perform One Step}
    	\label{Perform One Step}
    	\begin{algorithmic}[1]
    		\Procedure{PerformOneStep}{graphs, temperatures, cooling\_rates}
    		\State $\lambda$\_list = $[]$
    	    \For {k in 1: length(graphs)}    	    
    	    \State $G0 = $graphs$[k]$
    	    \State $\tau0$ = temperatures$[k]$
    	    \State $T$\_decay = cooling\_rates$[k]$
    	    \State $G1$, $\lambda1$ $\tau1$ = PartialAnneal($G0$, $T = \tau0$, cooling\_rate = $T$\_decay, switches $= k$)
    	    \State $\lambda$\_list $\gets \lambda$\_list $+ \lambda1$
    	    \State temperatures$[k]$ = $\tau1$
    	    \State graphs$[k] = G1$
    	    \EndFor
    	    \State ReverseRankOrder = argsort$(\lambda$\_list$)$
    	    \State graphs = graphs[ReverseRankOrder]
    	    \State temperatures = temperatures[ReverseRankOrder]
    	    \Comment Now randomly swap a pair of graphs
    	    \If {length(graphs) $> 2$}
    	    \State randomLocations = sampleFrom([1:length(graphs)], number = 2, without replacement)
    	    \State swap(graphs[randomLocations[1]],graphs[randomLocations[2]])
    	    \EndIf
    	    \State return graphs, temperatures, minimum($\lambda$\_list)
    		\EndProcedure
    	\end{algorithmic}	
    \end{algorithm}   
    
    	\begin{algorithm}
    	\caption{Coupled Annealing}
    	\label{Coupled Annealing}
    	\begin{algorithmic}[1]
    		\Procedure{CoupledAnnealing}{vertices, degree, chains, min\_cooling, max\_cooling, $T$\_min}
    	\State graphs, $tau$s, cooling\_rates = DefineCoupling(vertices, degree, chains, min\_cooling, max\_cooling)
    	\State best\_$\lambda$ = computeSecondEigenvalue(graphs[1])
    	\State best\_$G$ = graphs[1]
    	\While {maximum($\tau$s) $> T$\_min}
    	\State graphs, $\tau$s, current\_$\lambda$ = performOneStep(graphs, $\tau$s, cooling\_rates)
    	\If {current\_$\lambda < $ best\_$\lambda$}
    	\State best\_$\lambda =$ current\_$\lambda$
    	\State best\_$G$ = graphs[1]
    	\EndIf
    	\EndWhile
    	\State return best\_$G$, best\_$\lambda$
    	\EndProcedure
    	\end{algorithmic}	
    \end{algorithm}

\end{document}